\documentclass[12pt, preprint]{aastex}

\shorttitle{Stellar activity and detectability of solar-like oscillations}
\shortauthors{Chaplin et al.}

\begin{document}

\author{
   W.~J.~Chaplin\altaffilmark{1},
   T.~R.~Bedding\altaffilmark{2},
   A.~Bonanno\altaffilmark{3},
   A.-M.~Broomhall\altaffilmark{1},
   R.~A.~Garc\'ia\altaffilmark{4},
   S.~Hekker\altaffilmark{1,5},
   D.~Huber\altaffilmark{2},
   G.~A.~Verner\altaffilmark{6},
   S.~Basu\altaffilmark{7},
   Y.~Elsworth\altaffilmark{1},
   G.~Houdek\altaffilmark{8},
   S.~Mathur\altaffilmark{9},
   B.~Mosser\altaffilmark{10},
   R.~New\altaffilmark{11},
   I.~R.~Stevens\altaffilmark{1},
   T.~Appourchaux\altaffilmark{12},
   C.~Karoff\altaffilmark{13},
   T.~S.~Metcalfe\altaffilmark{9},
   J.~Molenda-\.Zakowicz\altaffilmark{14},
   M.~J.~P.~F.~G.~Monteiro\altaffilmark{15},
   M.~J.~Thompson\altaffilmark{9},
   J.~Christensen-Dalsgaard\altaffilmark{13},
   R.~L.~Gilliland\altaffilmark{16},
   S.~D.~Kawaler\altaffilmark{17},
   H.~Kjeldsen\altaffilmark{13},
   J.~Ballot\altaffilmark{18},
   O.~Benomar\altaffilmark{12},
   E.~Corsaro\altaffilmark{3},
   T.~L.~Campante\altaffilmark{13,15},
   P.~Gaulme\altaffilmark{12},
   S.~J.~Hale\altaffilmark{1},
   R.~Handberg\altaffilmark{13},
   E.~Jarvis\altaffilmark{1},
   C.~R\'egulo\altaffilmark{19,20},
   I.~W.~Roxburgh\altaffilmark{6},
   D.~Salabert\altaffilmark{19,20},
   D.~Stello\altaffilmark{2},
   F.~Mullally\altaffilmark{21},
   J.~Li\altaffilmark{21},
   W.~Wohler\altaffilmark{22}
}
\altaffiltext{1}{School of Physics and Astronomy, University of Birmingham, Edgbaston, Birmingham, B15 2TT, UK}

\altaffiltext{2}{Sydney Institute for Astronomy (SIfA), School of Physics, University of Sydney, NSW 2006, Australia}

\altaffiltext{3}{INAF Osservatorio Astrofisico di Catania, Via S.Sofia 78, 95123, Catania, Italy}

\altaffiltext{4}{Laboratoire AIM, CEA/DSM -- CNRS -- Universit\'e Paris Diderot -- IRFU/SAp, 91191 Gif-sur-Yvette Cedex, France}

\altaffiltext{5}{Astronomical Institute, "Anton Pannekoek", University of Amsterdam, PO Box 94249, 1090 GE Amsterdam, The Netherlands}

\altaffiltext{6}{Astronomy Unit, Queen Mary, University of London, Mile End Road, London, E1 4NS, UK}

\altaffiltext{7}{Department of Astronomy, Yale University, P.O. Box 208101, New Haven, CT 06520-8101, USA}

\altaffiltext{8}{Institute of Astronomy, University of Vienna, A-1180 Vienna, Austria}

\altaffiltext{9}{High Altitude Observatory and, Scientific Computing Division, National Center for Atmospheric Research, Boulder, Colorado 80307, USA}

\altaffiltext{10}{LESIA, CNRS, Universit\'e Pierre et Marie Curie, Universit\'e Denis Diderot, Observatoire de Paris, 92195 Meudon cedex, France}

\altaffiltext{11}{Materials Engineering Research Institute, Faculty of Arts, Computing, Engineering and Sciences, Sheffield Hallam University, Sheffield, S1 1WB, UK}

\altaffiltext{12}{Institut d'Astrophysique Spatiale, Universit\'e Paris XI -- CNRS (UMR8617), Batiment 121, 91405 Orsay Cedex, France}

\altaffiltext{13}{Department of Physics and Astronomy, Aarhus University, DK-8000 Aarhus C, Denmark}

\altaffiltext{14}{Astronomical Institute, University of Wroc\l{}aw, ul. Kopernika, 11, 51-622 Wroc\l{}aw, Poland}

\altaffiltext{15}{Centro de Astrof\'\i sica and Faculdade de Ci\^encias, Universidade do Porto, Rua das Estrelas, 4150-762, Portugal}

\altaffiltext{16}{Space Telescope Science Institute, Baltimore, MD 21218, USA}

\altaffiltext{17}{Department of Physics and Astronomy, Iowa State University, Ames, IA 50011, USA}

\altaffiltext{18}{Institut de Recherche en Astrophysique et
Plan\'etologie, Universit\'e de Toulouse, CNRS, 14 av E. Belin, 31400
Toulouse, France}

\altaffiltext{19}{Departamento de Astrof\'{\i}sica, Universidad de La Laguna, E-38206 La Laguna, Tenerife, Spain}

\altaffiltext{20}{Instituto de Astrof\'{\i}sica de Canarias, E-38200 La Laguna, Tenerife, Spain}

\altaffiltext{21}{SETI Institute/NASA Ames Research Center, Moffett Field, CA 94035, USA}

\altaffiltext{22}{Orbital Sciences Corporation/NASA Ames Research Center, Moffett Field, CA 94035, USA}

\title{Evidence for the impact of stellar activity on the detectability of solar-like oscillations observed by \emph{Kepler}}

\begin{abstract}

We use photometric observations of solar-type stars, made by the NASA
\emph{Kepler Mission}, to conduct a statistical study of the impact of
stellar surface activity on the detectability of solar-like
oscillations. We find that the number of stars with detected
oscillations fall significantly with increasing levels of
activity. The results present strong evidence for the impact of
magnetic activity on the properties of near-surface convection in the
stars, which appears to inhibit the amplitudes of the stochastically
excited, intrinsically damped solar-like oscillations.

\end{abstract}


\section{Introduction}
\label{sec:intro}

Solar-type stars show ``solar-like'' acoustic oscillations that are
intrinsically damped and stochastically excited by near-surface
convection (e.g., Houdek et al. 1999; Christensen-Dalsgaard 2004;
Samadi et al. 2007).  It is now well established that magnetic
structures in the solar photosphere are strong absorbers of acoustic
(or $p$-mode) oscillations (Braun et al. 1987, 1988; Braun \& Birch
2008). A strong magnetic field can diminish the turbulent velocities
in a convectively unstable layer (e.g., Proctor \& Weiss 1982;
Cattaneo et al. 2003) and this can affect the driving of acoustic
modes (e.g., Jacoutot et al. 2008).  In solar-type stars, the presence
of a fibral magnetic field (Gough \& Thompson 1988; Goldreich et
al. 1991; Houdek et al. 2001) may become sufficiently strong to affect
not only the properties of the $p$-mode propagation, but also the
turbulence of the convection by reducing its magnitude with increasing
stellar activity, thereby reducing the amplitudes of the oscillations.

The amplitudes (i.e., the square root of the total powers) of solar
$p$ modes are observed to decrease with increasing levels of solar
activity (Chaplin et al. 2000; Komm et al. 2000). The decrease
observed from solar minimum to solar maximum is about 12.5\,\% for
modes of low spherical degree, $l$ (Chaplin et al. 2000; Gelly et
al. 2002; Jim\'enez-Reyes et al. 2003), which are the modes that are
detectable in observations of solar-type stars.  Garc\'ia et
al. (2010) recently uncovered the first evidence for changes in
$p$-mode amplitudes associated with a stellar activity cycle in
another solar-type star, from \emph{CoRoT} satellite data on HD49933.

HD49933, and the other F-type stars observed for asteroseismology by
\emph{CoRoT}, have activity levels that are not disimilar to those of
the active Sun, as discerned from levels of variability in the
lightcurves arising from rotational modulation of starspots and active
regions (Mosser et al. 2005; 2009a). The same is true for the F5 star
Procyon, as measured in both radial velocity (Arentoft et al. 2008)
and photometry (Huber et al., 2010). However, the G-type dwarf
HD175726, which was also observed by \emph{CoRoT} (Mosser et
al. 2009b), shows much higher levels of activity but barely detectable
solar-like oscillations. Mosser et al. speculated that the
lower-than-expected amplitudes might have resulted from suppression by
high levels of intrinsic magnetic activity. Dall et al. (2010) made a
simlar suggestion for the active G8 star EK Eri. Here, we use the
unprecedented large ensemble of oscillating solar-type stars observed
by the NASA \emph{Kepler Mission} to search for evidence of this
effect in a large number of stars.

In addition to searching for exoplanets (Borucki et al. 2010; Koch et
al. 2010), \emph{Kepler} is providing large quantities of high-quality
data for the asteroseismic investigation of stars, as part of the
\emph{Kepler} Asteroseismology Investigation (Gilliland et
al. 2010a). Photometry of a subset of these stars is being made at a
cadence that is rapid enough to allow investigations of oscillations
in solar-type stars, where dominant periods are of the order of
several minutes (Chaplin et al. 2010, 2011a; Christensen-Dalsgaard et
al. 2010; Metcalfe et al. 2010). During the first seven months of
science operations just over 2000 stars were observed for one month
each as part of an asteroseismic survey of the solar-type part of the
color-magnitude diagram. Solar-like oscillations have been detected in
about 500 stars, increasing by a factor of about 25 the number of
solar-type stars with detected oscillations.

The large number of solar-type stars in this \emph{Kepler} ensemble
makes possible the statistical study of intrinsic stellar properties
and trends, in what is a homogenous data sample of unprecedented
quality. Here, we use results on the ensemble to conduct a statistical
study of the impact of stellar activity on the detectability of
solar-like oscillations.

For this study we have used two simple measures of variability in the
\emph{Kepler} lightcurves as proxies of the levels of stellar magnetic
activity, including one measure suggested by Basri et al. (2010, 2011)
from their survey of activity levels in more than 100,000 stars
observed in 30-min (long) cadence by \emph{Kepler} during its first
month of science operations. We study a subset of about 2000
solar-type stars having short-cadence \emph{Kepler} data and test if
the distribution of observed variability differs for stars that do,
and do not, have detected solar-like oscillations.

\section{Data and Analysis}
\label{sec:anal}

We use asteroseismic results on solar-type stars that were observed by
\emph{Kepler} during the first seven months of science
operations. About 2000 stars, down to \emph{Kepler} apparent magnitude
$Kp \simeq 12.5$, were selected as potential solar-type targets based
upon parameters in the \emph{Kepler} Input Catalog (KIC; Batalha et
al. 2010, Koch et al. 2010). Each star was observed for one month at a
time in short-cadence mode (58.85\,s; see Gilliland et
al. 2010b). Time series were prepared for asteroseismic analysis in
the manner described by Garc\'ia et al. (2011), using procedures that
work on the raw lightcurves. Lightcurves prepared for the Transiting
Planet Search pipeline are not appropriate for use here since
phenomena associated with stellar variability are suppressed or
removed to aid planet detection (Jenkins et al. 2010).

Different teams analyzed the prepared lightcurves to attempt to detect
signatures of solar-like oscillations (see Huber et al. 2009; Mosser
\& Appourchaux 2009; Roxburgh 2009; Campante et al. 2010; Chaplin et
al. 2010, 2011b; Hekker et al. 2010; Karoff et al. 2010; Mathur et
al. 2010). Most of the applied detection techniques relied on
extracting signatures of the near-regular frequency separations of the
solar-like oscillation frequency spectrum, while others searched for
signatures of the Gaussian-like power excess due to the
oscillations. Verner et al. (2011) present a detailed comparison of
the results returned by the different pipelines on the \emph{Kepler}
lightcurves, finding reassuring levels of agreement between the
results.  We demanded that at least two of the asteroseismic data
analysis pipelines returned consistent results on a star for it to be
flagged as a solid detection for use in this paper. A total of around
500 stars were flagged as having detected solar-like oscillations. The
signal-to-noise ratio in the oscillations required for a detection was
$\ga 0.1$, as defined by the ratio of the maximum power spectral
density of the smoothed oscillation envelope relative to the estimated
background at the frequency of maximum oscillations power. This
threshold level allows an unambigous detection of the signature of the
large frequency separation, and identification of a significant power
excess due to the oscillations.

For each star that was analyzed, we used two simple metrics of
variability as proxies of the stellar surface activity, as determined
from direct analysis of the prepared lightcurves. For both, we first
applied a low-pass filter to remove the oscillation signal smoothing
each lightcurve with a one-hr-long boxcar. For one metric, we measured
the maximum absolute deviation of each smoothed lightcurve from its
mean. This simple measure of variability is what Basri et al. (2010)
call the ``range'', and here we label it $r_{\rm hr}$. Basri et
al. smoothed long-cadence (29.4\,min) \emph{Kepler} lightcurves with a
10-hr boxcar. We tested the effect of smoothing our short-cadence
lightcurves on timescales ranging from 1\,hr up to 1\,d, but found no
significant impact on our results. As our second variability metric we
use what Garc\'ia et al. (2010) referred to as a ``starspot
proxy''. We measured the standard deviation about the mean (i.e., the
\textsc{rms}) of each smoothed lightcurve, and we call this metric
$\sigma_{\rm hr}$.

Neither of the measures, as defined above, takes explicit account of
the apparent magnitude of the target. We would expect there to be a
magnitude-dependent correction to $r_{\rm hr}$ and $\sigma_{\rm hr}$
on account of the changing contribution due to shot noise, although
the effect is small. We specified the required corrections using the
``minimal noise'' model for \emph{Kepler} in Gilliland et
al. (2010b). The \textsc{rms} noise, $\sigma$, per $\Delta t =
58.85$-sec integration is given by
   \begin{equation}
   \sigma = \frac{10^3}{c} \left( c+9.5 \times 10^5(14/Kp)^5
   \right)^{1/2}\,\rm ppt,
   \label{eq:noise}
   \end{equation}
where $c = 1.28 \times 10^{0.4(12-Kp)+7}$ detections~per~cadence, and
$Kp$ is the \emph{Kepler} apparent magnitude. Since the time series
are smoothed with a 1-hr (3600-sec) boxcar, the additive correction
that must be removed from $\sigma_{\rm hr}$ is just $\sigma(\Delta
t/3600)^{1/2}$. The correction for $r_{\rm hr}$ was calibrated with
Monte-Carlo simulations, and found to be $\sim (2/3)\sigma$.

Since our aim is to understand the impact of stellar activity on the
detectability of the solar-like oscillations, we ignored those stars
whose lightcurve variability could be attributed to another
phenomenon, e.g., eclipsing binaries, and classical pulsators at the
hot end of the sample. The fraction of eclipsing binaries that we
removed ($\sim 1\,\%$) is in line with the rate of occurence of
binaries in the \emph{Kepler} field of view, as determined by Pr\u{s}a
et al. (2011) down to $Kp \sim 16$.

\section{Results}
\label{sec:res}

\begin{figure*}
\epsscale{1.1}
\plottwo{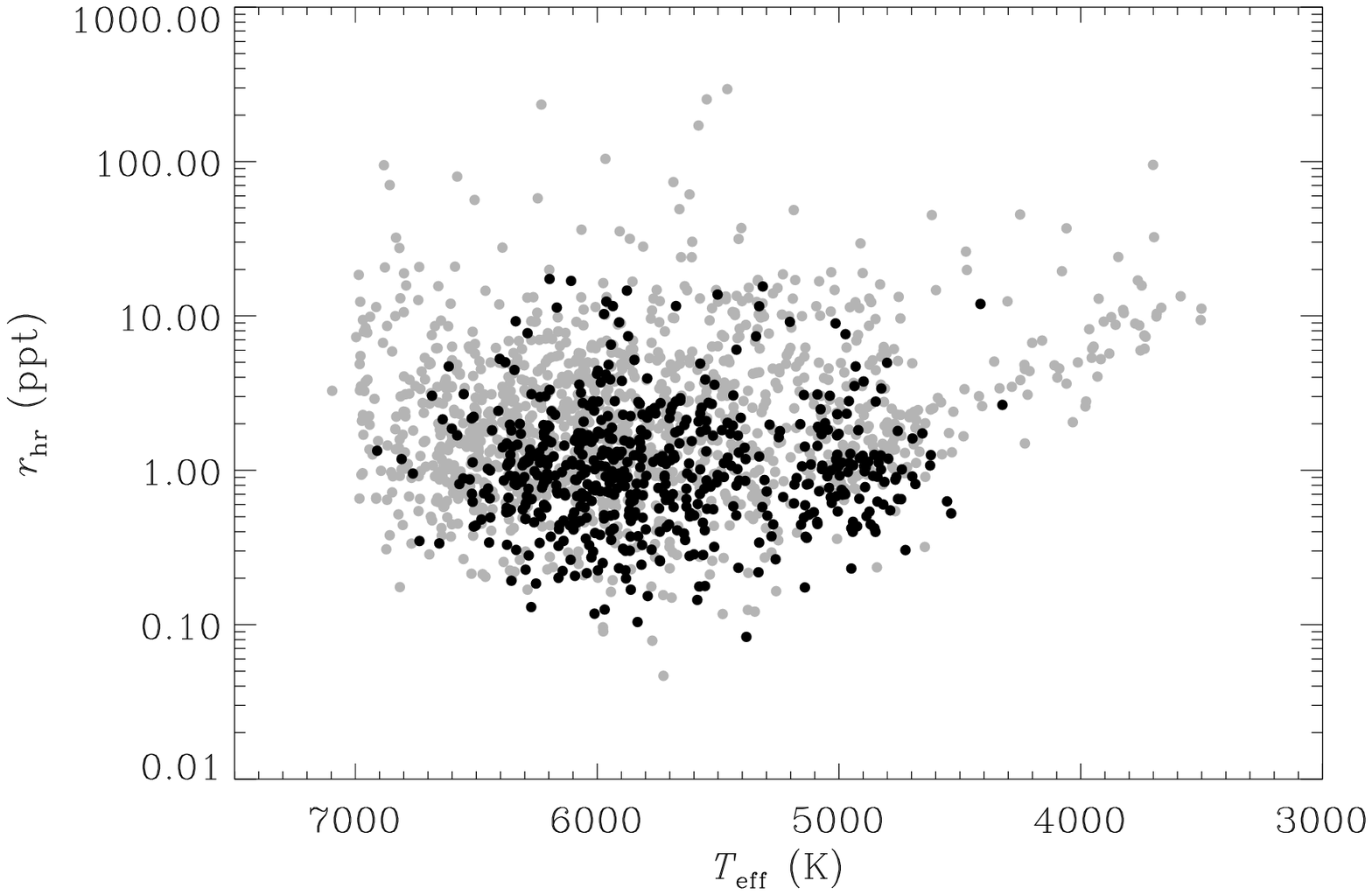}{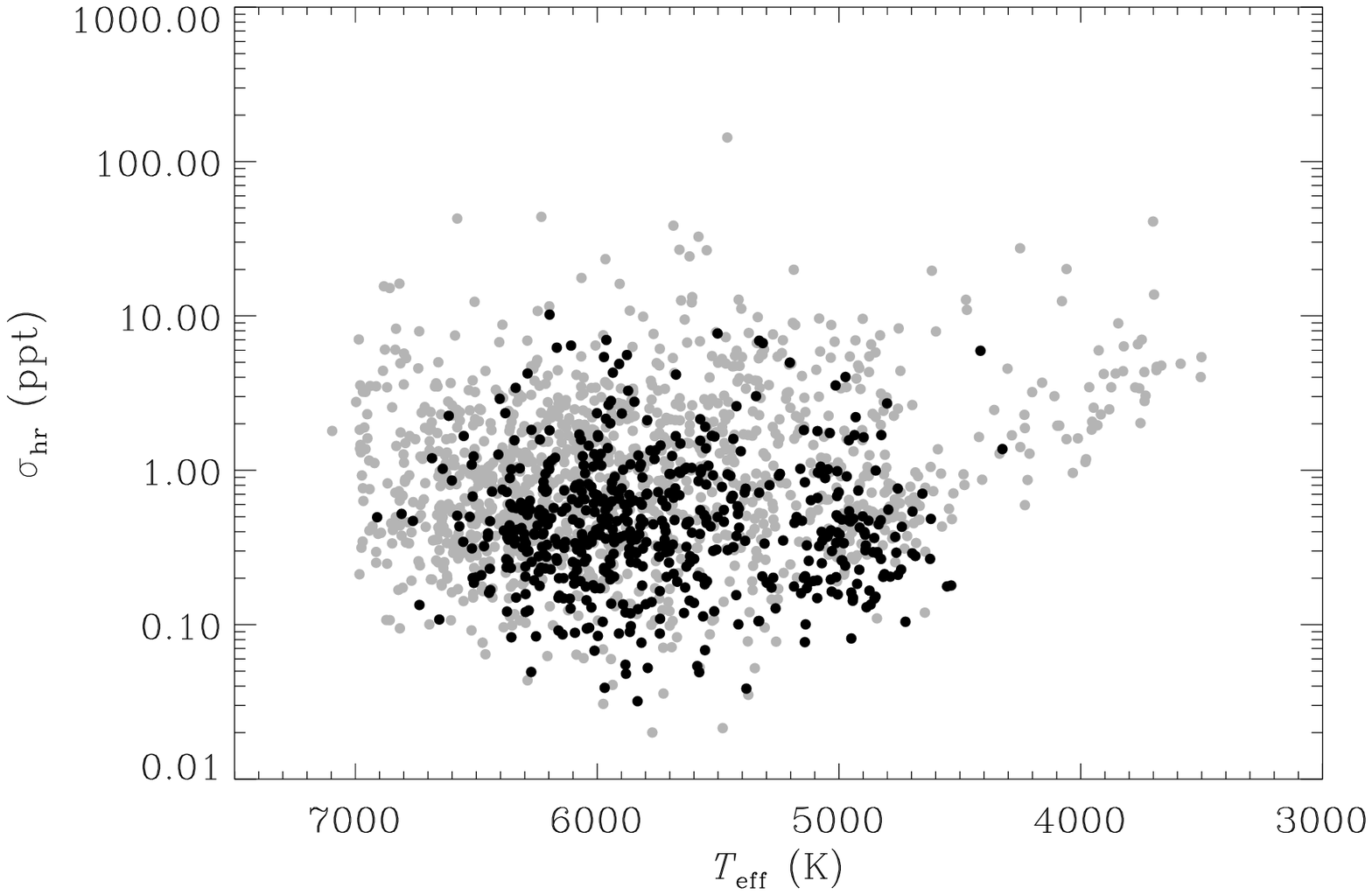}
\plottwo{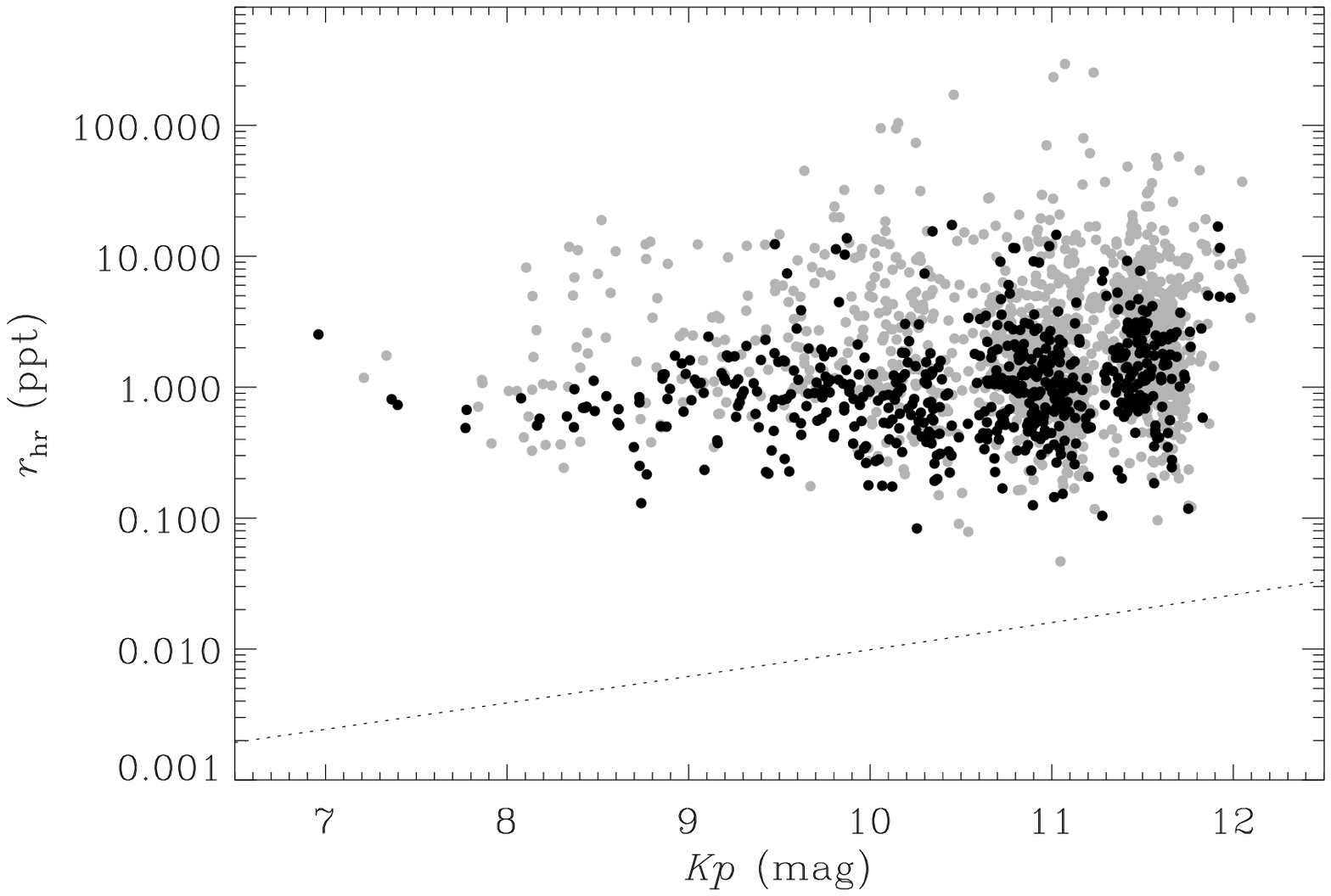}{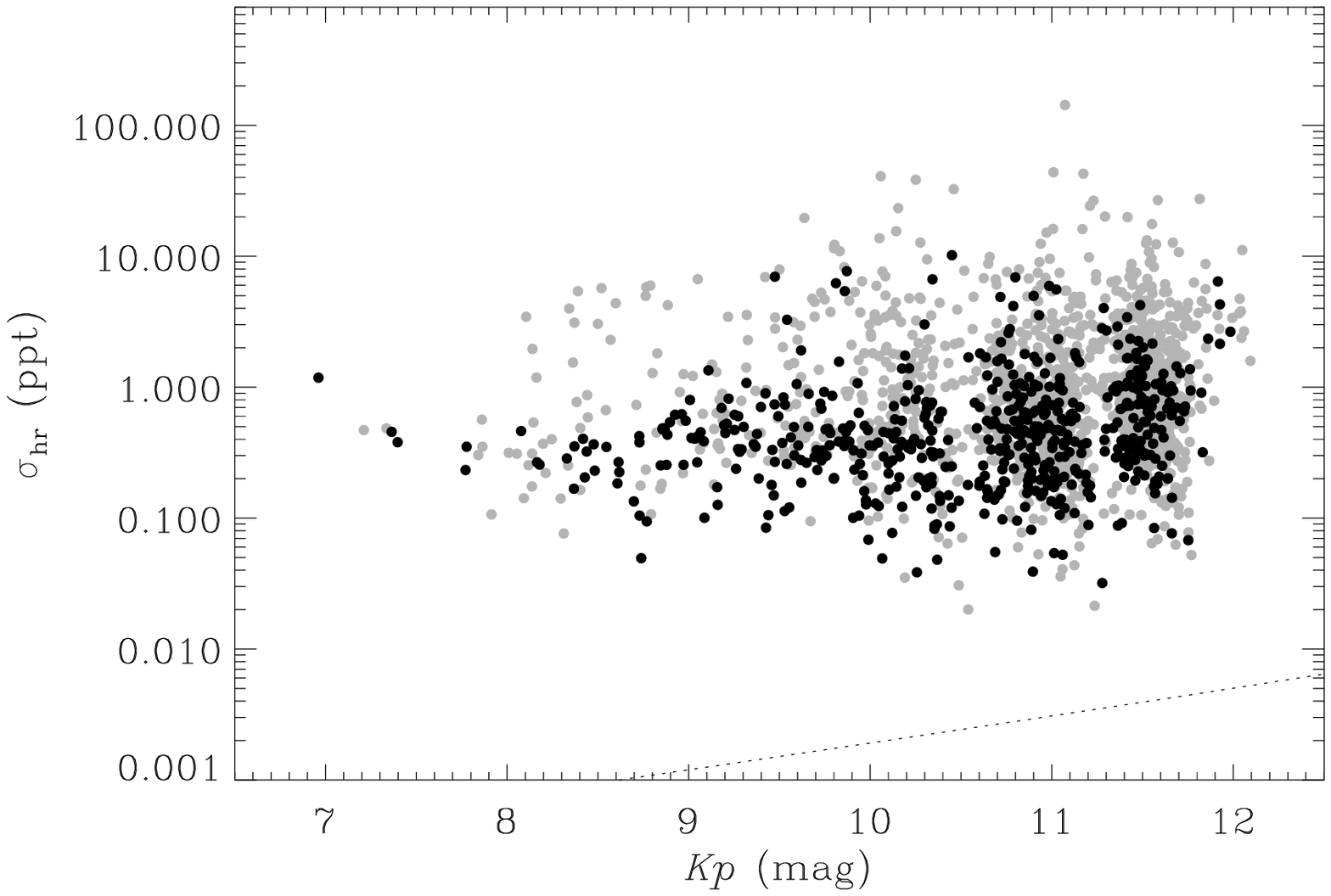}
\caption{Range, $r_{\rm hr}$, and \textsc{rms}, $\sigma_{\rm hr}$ as a
function of $T_{\rm eff}$ (top panels) and \emph{Kepler} apparent
magnitude, $Kp$ (bottom panels). Stars with detected solar-like
oscillations are plotted in black; stars with no detections are
plotted in gray. The dotted lines follow the additive corrections that
were applied to $r_{\rm hr}$ and $\sigma_{\rm hr}$ (see text).}
\label{fig:metric}
\end{figure*}

The top two panels of Fig.~\ref{fig:metric} plot the range, $r_{\rm
hr}$, and the \textsc{rms}, $\sigma_{\rm hr}$, as a function of
$T_{\rm eff}$. Points in black are stars with detected solar-like
oscillations. The bottom two panels plot the metrics with the
\emph{Kepler} apparent magnitude, $Kp$, as the independent
variable. The dotted lines follow the additive corrections (see
above).

%%%%%%%%%%%%%%%%%%%%%%%%%%%%%%%%%%%%%%%%%%%%%%%%%%%%%%%%%%%%%%%%%%%%%%%

%%%%%%%%%%%%%%%%%%%%%%%%%%%%%%%%%%%%%%%%%%%%%%%%%%%%%%%%%%%%%%%%%%%%%%%

\begin{figure*}
\epsscale{1.1}
\plottwo{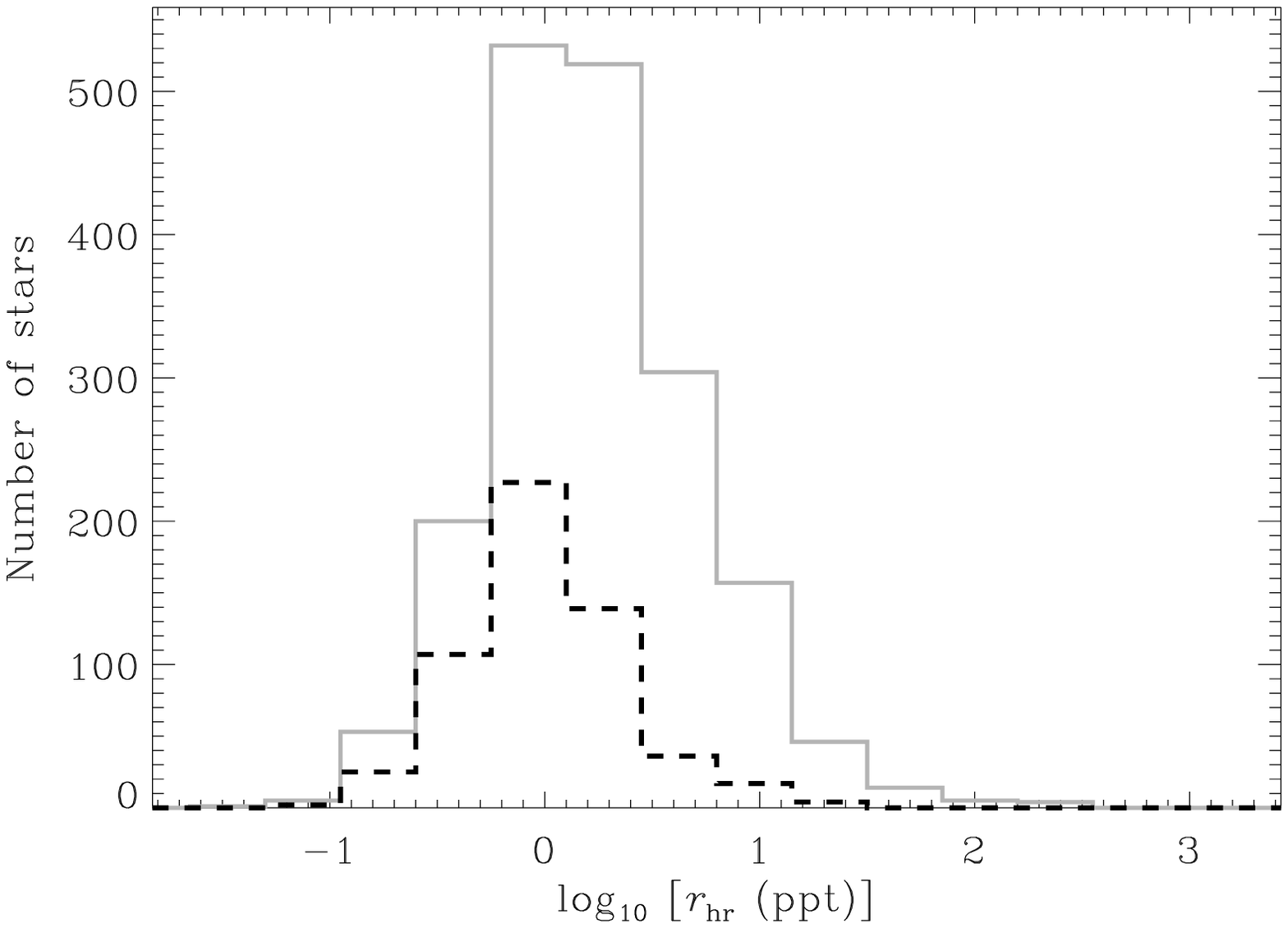}{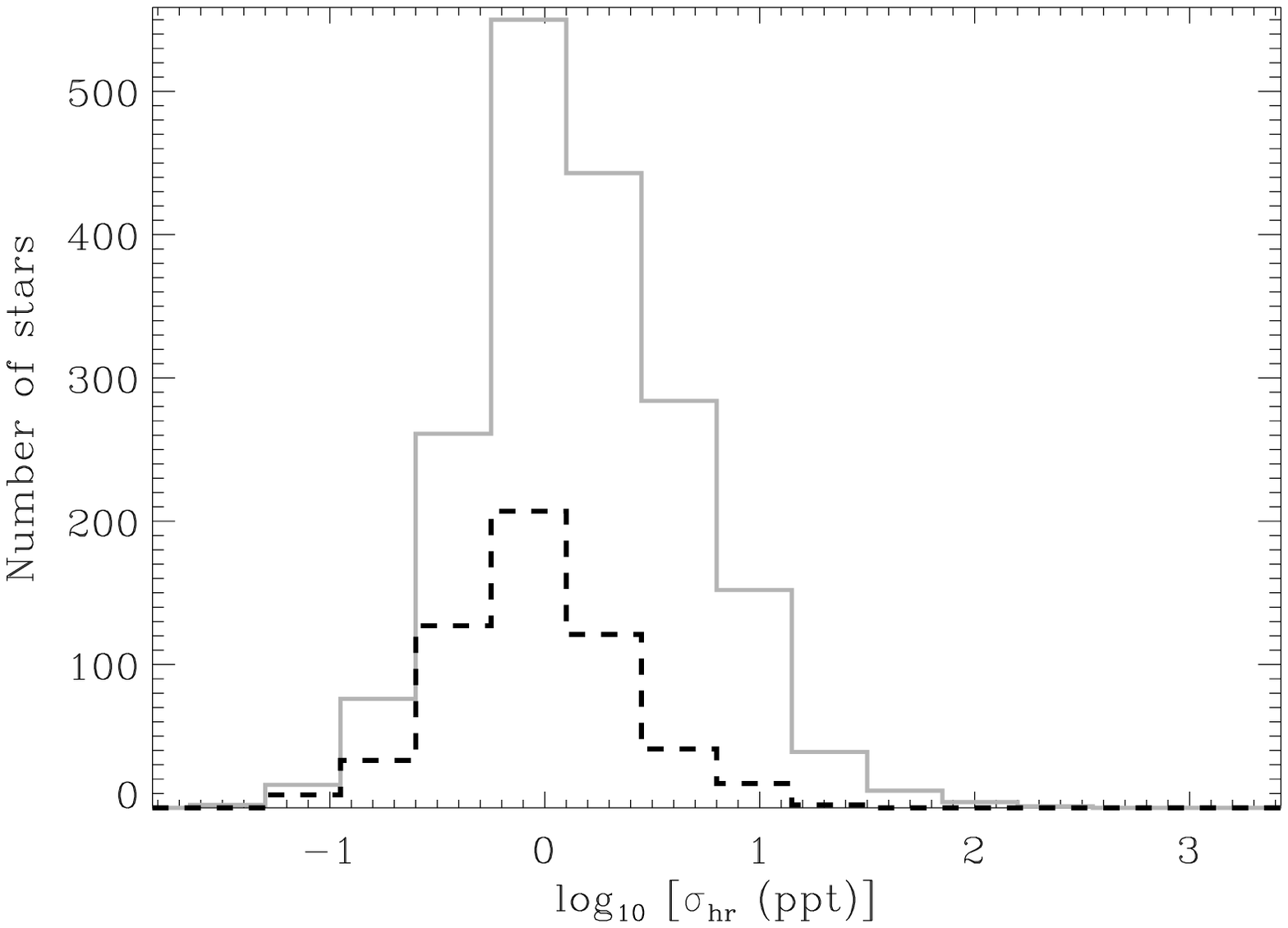}
\plottwo{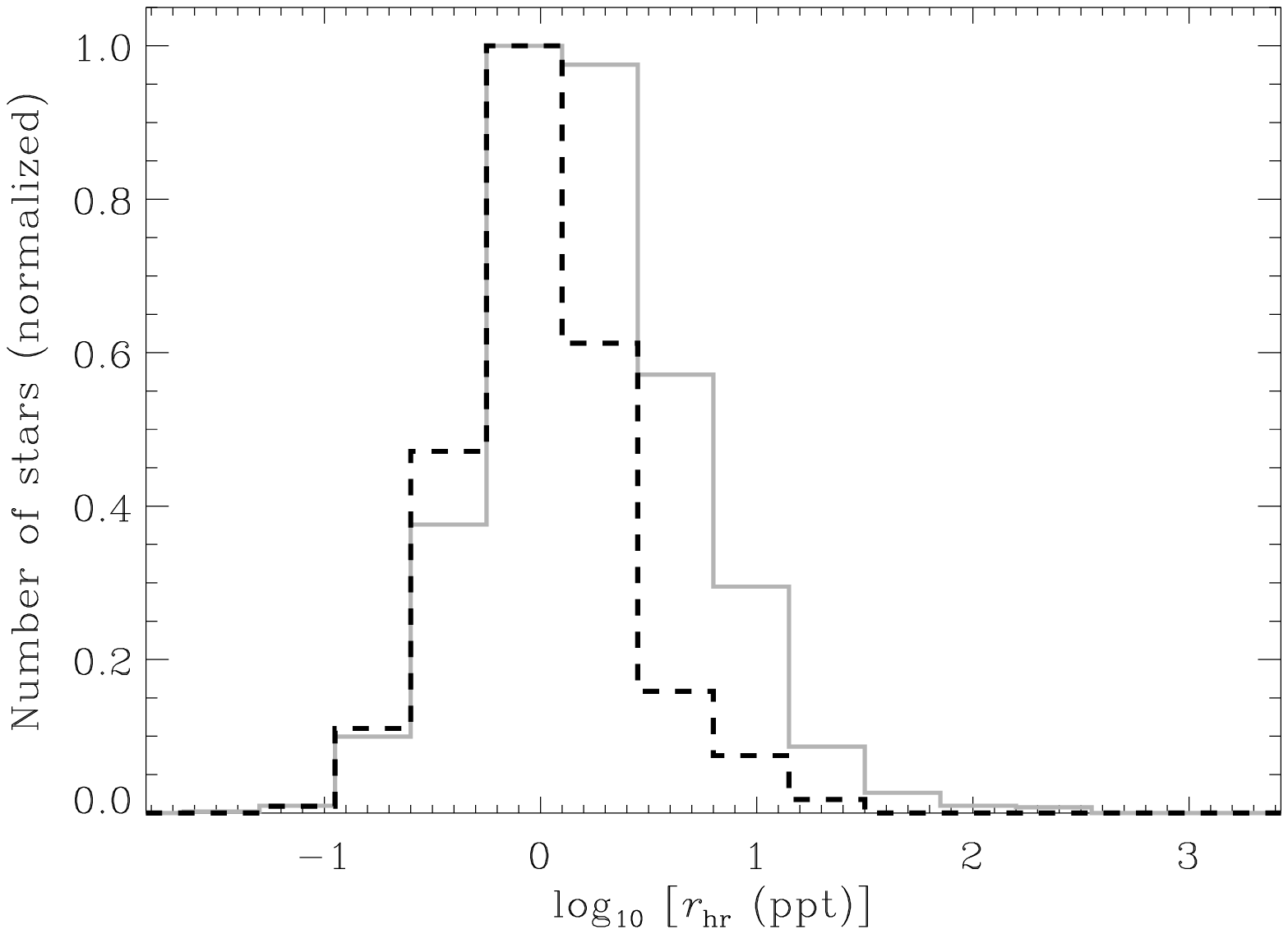}{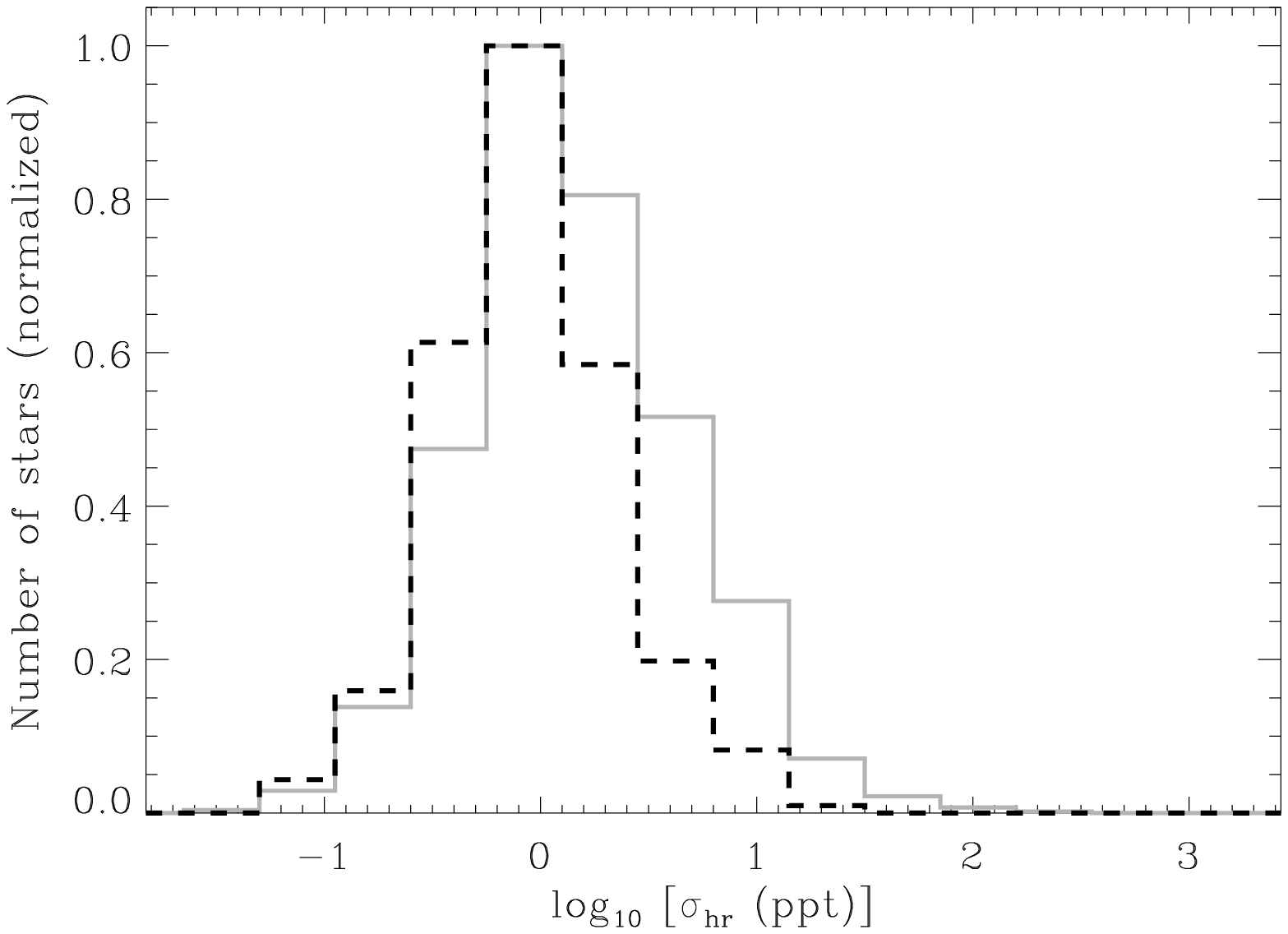}
\plottwo{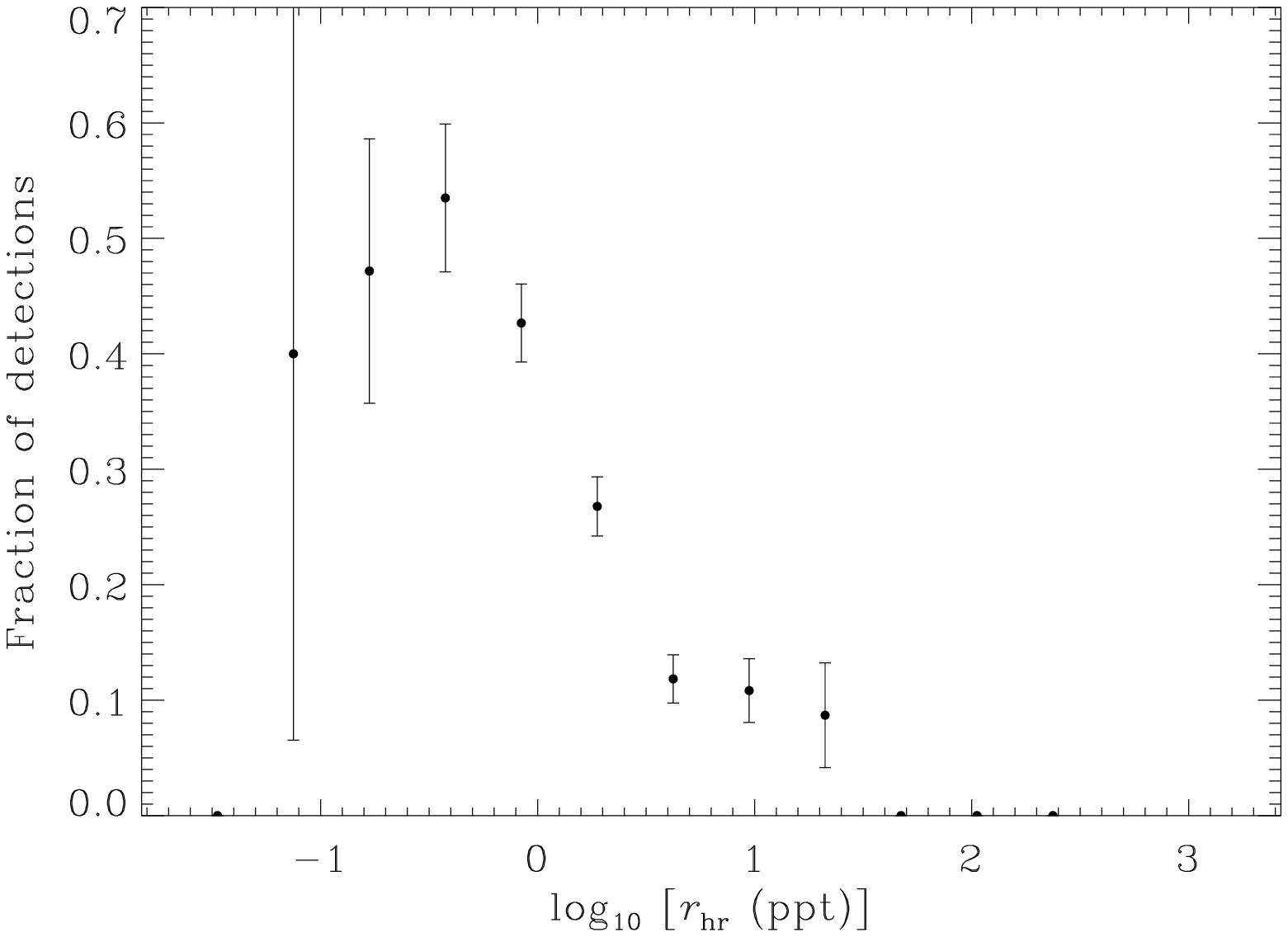}{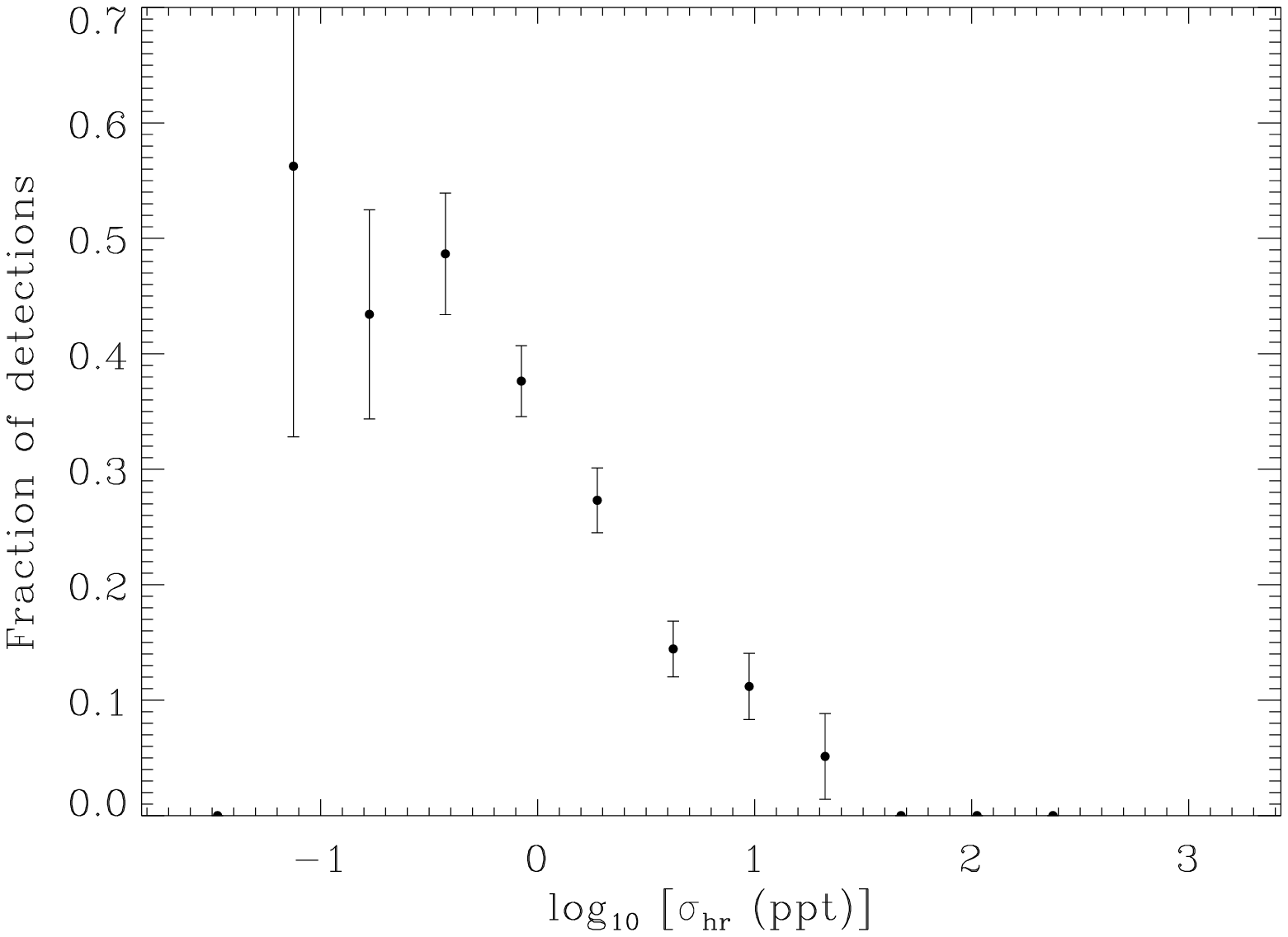}
\caption{Top panels: Histograms of $r_{\rm hr}$ and $\sigma_{\rm hr}$,
  for all analyzed stars (gray solid lines) and stars with detected
  solar-like oscillations (black dashed lines). Middle panels:
  Histograms normalized to a maximum value of unity (same
  linestyles). Bottom panels: Fraction of stars in each histogram bin
  showing detected solar-like oscillations.}
\label{fig:hist}
\end{figure*}

%%%%%%%%%%%%%%%%%%%%%%%%%%%%%%%%%%%%%%%%%%%%%%%%%%%%%%%%%%%%%%%%%%%%%%%

Our range plot in Fig.~\ref{fig:metric} is very similar to the
corresponding plot in Basri et al. (2010), which shows results on just
over 100,000 stars observed during the first full month of
\emph{Kepler} science operations.  The data have a lower-limit
envelope that has its minimum at approximately solar temperature. The
lower limit of the envelope shifts to higher levels of variability at
lower and higher temperatures.

It is also apparent from Fig.~\ref{fig:metric} that the cloud of
points for stars \emph{with} detected solar-like oscillations (black
points) is clustered toward lower levels of variability. We see no
detections above $r_{\rm hr}$ of $\approx 20\,\rm ppt$, and
$\sigma_{\rm hr}$ of $\approx 10\,\rm ppt$.  The cloud of points for
stars \emph{without} detected oscillations (gray points) extends to
much higher values in both $r_{\rm hr}$ and $\sigma_{\rm hr}$.  There are
also stars having no detections that nevertheless show the same levels
of variability as stars with detections, i.e., at the lower levels of
variability the distributions overlap. The apparent gap at $T_{\rm
eff} \simeq 5300\,\rm K$ in the distribution of stars with detected
oscillations is the result of higher numbers of detections at lower
$T_{\rm eff}$ from evolved stars at the base of the red giant branch,
which have higher oscillation amplitudes than their main-sequence
cousins. The distributions in $Kp$ (bottom panels of
Fig.~\ref{fig:metric}) are also not smooth, but this effect is present
for stars both with, and without, detected oscillations, and it is
merely a selection effect arising from the final choice of target
stars.

Histograms of $r_{\rm hr}$ and $\sigma_{\rm hr}$ are plotted in the
top panels of Fig.~\ref{fig:hist}, for all analyzed stars (gray solid
lines) and stars with detected oscillations only (black dashed
lines). The middle panels show the histograms normalized to a maximum
value of unity to allow a visual comparison to be made of the shapes
of the distributions of stars with, and without, detected
oscillations. The bottom panels plot the fraction of stars in each
histogram bin that show detections, with the errors calculated from
Poisson statistics. The histograms in the middle panels show the
aforementioned significant deficit of detections at higher levels of
variability. The significance of this fall-off is confirmed by the
detection ratios in the bottom panels.

Two categories of explanation for the fall-off suggest themselves: one
that has its origins in the intrinsic properties of the stars (much
the more interesting explanation); or another that is the result of
either data-analysis or selection bias issues affecting the
detectability of the modes. Let us consider first the possibility that
data analysis issues might be the cause.

Sudden discontinuities or large excursions in the lightcurves can
affect the appearance of the frequency spectrum of the lightcurves if
not treated properly, e.g., as a result of the introduction of a
complex overtone structure, or the introduction of frequency-dependent
noise, into the frequency range of interest for detection of the
oscillations. To test the impact of these effects on our results we
selected about 150 stars for which we had successfully detected
oscillations, covering a wide range of intrinsic stellar properties
and apparent magnitudes.  We then picked a subset of 50 stars with
high levels of variability, but no detected oscillations. In one set
of tests -- analyzed using the pipeline described by Hekker et
al. (2010) -- we added the low-pass filtered parts of the high
variability lightcurves to the high-pass filtered parts of the
lightcurves with detected oscillations, and then checked whether we
could still detect the oscillations. In another set of tests --
analyzed using the pipeline described by Huber et al. (2009) -- we
selected high variability stars at random, and generated synthetic
lightcurves comprised of 10 random low-frequency sinusoids with
amplitudes selected to ensure that the low-frequency spectra of the
synthetic lightcurves matched those of the real stars. These synthetic
lightcurves were then multiplied by a random factor -- to allow as
wide a range of variability to be sampled as possible -- and then
added to one of the selected lightcurves with detected
oscillations. Again, we tested to see whether the oscillations could
still be detected.  The data-analysis pipelines detected oscillations
at all levels of variability (i.e., even in the most extreme cases),
and while a modest fall-off of the detection rates was seen at the
highest variability (e.g., for $r_{\rm hr} \ga 30\,\rm ppt$) this was
at nothing like the levels seen in the real \emph{Kepler} data.

Another concern would be that the fall-off is the result of selection
bias: for example, that stars showing higher levels of variability
also tend to be fainter, on the average, and therefore less likely to
show detected oscillations due to higher levels of shot noise. There
are certainly more stars at fainter $Kp$, which we would expect to
more fully sample the underlying distribution of $r_{\rm hr}$ and
$\sigma_{\rm hr}$, including the higher values. The bottom panels of
Fig.~\ref{fig:metric} do seem to bear this out. To test the possible
effects of selection bias, we ran the detection prediction code
developed for use by the \emph{Kepler} Science Team (Chaplin et
al. 2011b) on all 2000 targets. This code takes as input the KIC
radius, effective temperature and apparent magnitude, and produces as
output an estimate of the probability of detection for an assumed
length of observation, based on use of scaling relations that know
nothing about the possible effects of activity on the amplitudes of
the oscillatons.  We are interested only in the distribution of
predicted detections, and the ensemble is large enough to provide
statistically robust results. Those results indicated that the steep
fall-off seen in the \emph{Kepler} results cannot be explained by
selection bias. There was some fall-off in the predicted fraction of
detected oscillations, because there are more stars with high
variability that are also faint (see above); but we saw ``predicted''
detections in stars showing even the highest levels of
variability. The results also provided an explanation of why some
stars with low levels of variability did not have detected
oscillations. The expected success rate of detections was found to be
less than 100\,\%, because noise realizations will in some cases
hamper extraction of the oscillation signals.  That said, the absence
of detections in some of the brightest stars in the sample that show
only modest levels of variability is a puzzle. It may be that the
inclinations of stars play a role here.  The inclination will affect
the apparent (i.e., observed) variations seen in $r_{\rm hr}$ and
$\sigma_{\rm hr}$, assuming that in solar-type stars those variations
are dominated by contributions from active latitudes like for the Sun
(e.g., see Knaack et al. 2001; V\'azquez Ram\'io et al. 2011). It
could be that some stars with small $r_{\rm hr}$ and $\sigma_{\rm hr}$
are actually intrinsically active stars observed at low angles of
inclination (which will reduce the variations observed in the
lightcurves).

In summary, we conclude that the steep fall-off in the observed
fraction of detections has a stellar explanation.

\section{Discussion}
\label{sec:disc}

\begin{figure*}
\epsscale{1.1}
\plottwo{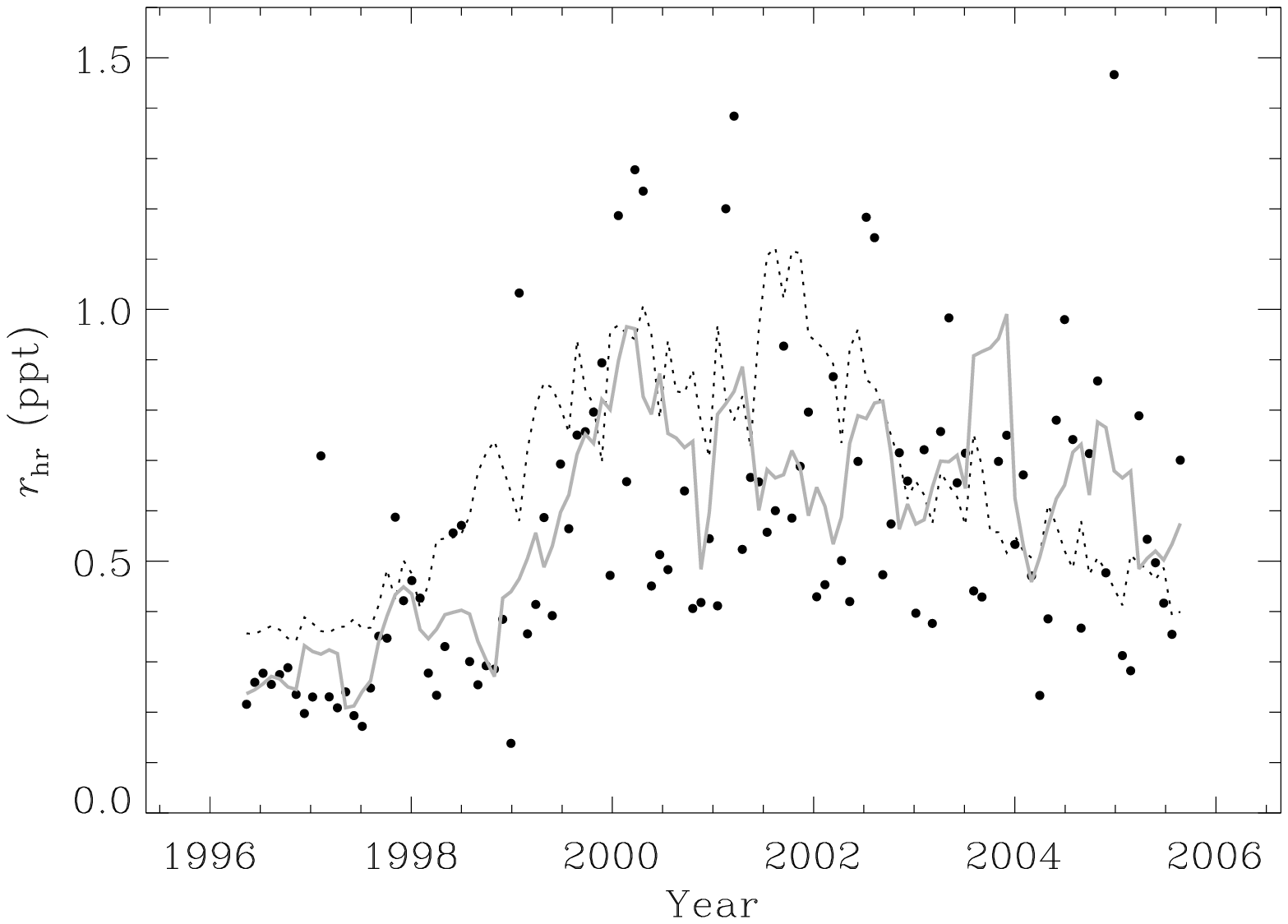}{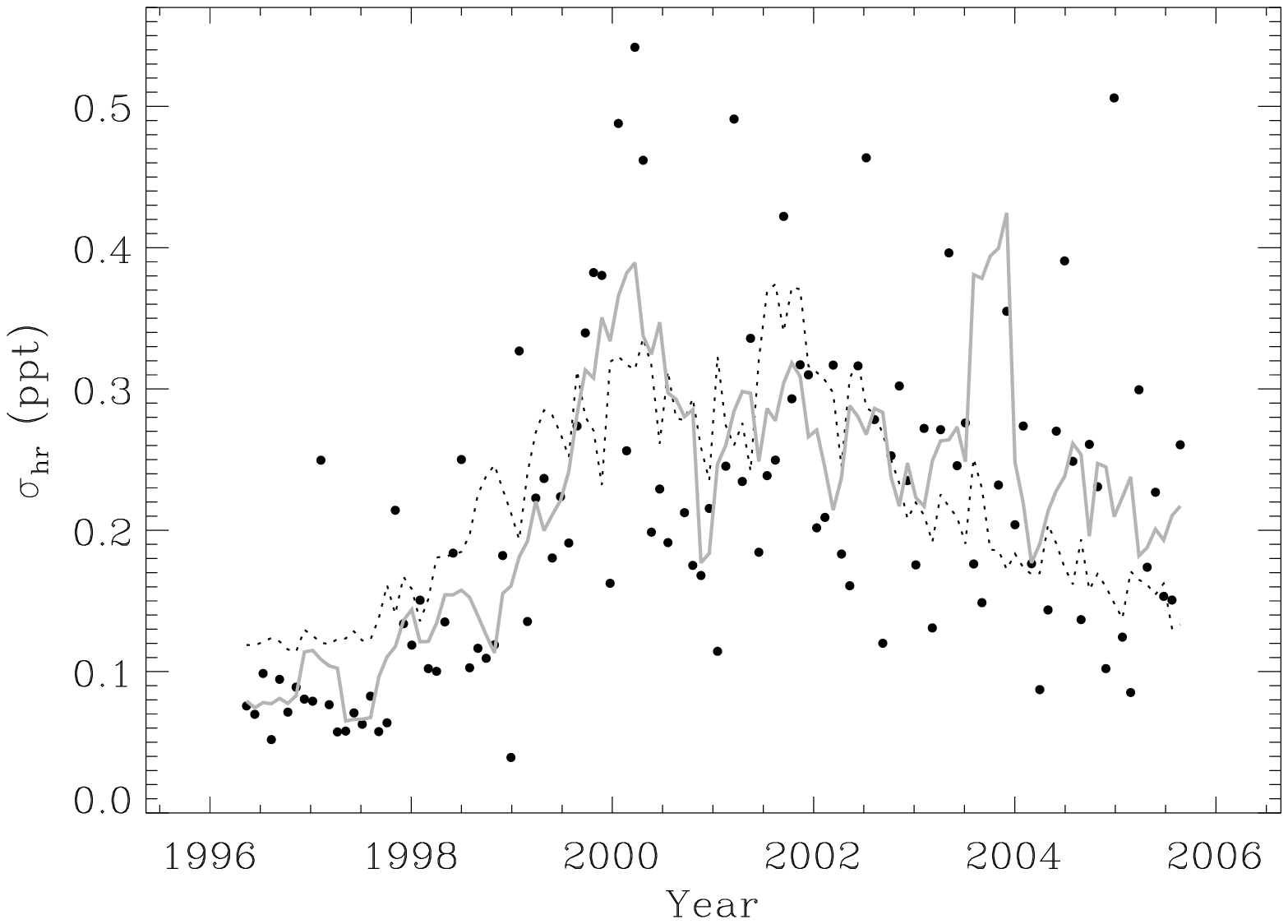}

\caption{Estimates of $r_{\rm hr}$ (left-hand panel) and $\sigma_{\rm hr}$
(right-hand panel) for the Sun (filled symbols), as determined from
analysis of one-month-long segments of PMO6 data. Gray lines describe
smooth curves through the independent measures, while the dotted lines
show the scaled 10.7-cm radio flux.}

\label{fig:sun}
\end{figure*}

The most compelling explanation for the results is that they show
evidence for intrinsic stellar (magnetic) activity suppressing the
amplitudes of the solar-like oscillations, and hence adversely
affecting the detectability of those oscillations.  In offering this
explanation we assume that the metrics we have used are reasonable
proxies of intrinsic levels of stellar activity (see Basri et
al. (2010) for further discussion). This certainly seems to be the
case for the Sun. We analyzed observations of the bolometric flux of
the Sun made by the PMO6 instrument onboard the \emph{ESA/NASA SOHO}
spacecraft (Fr\"ohlich et al. 1997). Fig.~\ref{fig:sun} plots $r_{\rm
hr}$ and $\sigma_{\rm hr}$ (filled symbols in each panel) as
determined by analysis of independent one-month-long segments of PMO6
data. The gray line in each panel describes a smooth curve through the
independent measures, while the dotted line shows the scaled 10.7-cm
radio flux, which is an excellent proxy of the global level of
magnetic activity on the Sun (e.g., see Chaplin et al. 2007, and
references therein). The underlying trends in $r_{\rm hr}$ and
$\sigma_{\rm hr}$ clearly follow those in the magnetic activity.

%%%%%%%%%%%%%%%%%%%%%%%%%%%%%%%%%%%%%%%%%%%%%%%%%%%%%%%%%%%%%%%%%%%%%%%

%%%%%%%%%%%%%%%%%%%%%%%%%%%%%%%%%%%%%%%%%%%%%%%%%%%%%%%%%%%%%%%%%%%%%%%

The range varies from about 0.25\,ppt to about 1\,ppt between solar
minimum and maximum. We know that the amplitudes of the low-degree
solar $p$ modes are at the same time suppressed by a fraction $\simeq
0.125$ (Chaplin et al. 2000; Gelly et al. 2002; Jim\'enez-Reyes et
al. 2003; Garc\'ia et al. 2010).  These solar data in principle allow
us to calibrate the expected suppression of oscillation amplitudes due
to activity. Here we present a simple estimate for $r_{\rm hr}$.

\begin{figure}
\epsscale{1.1}
\plotone{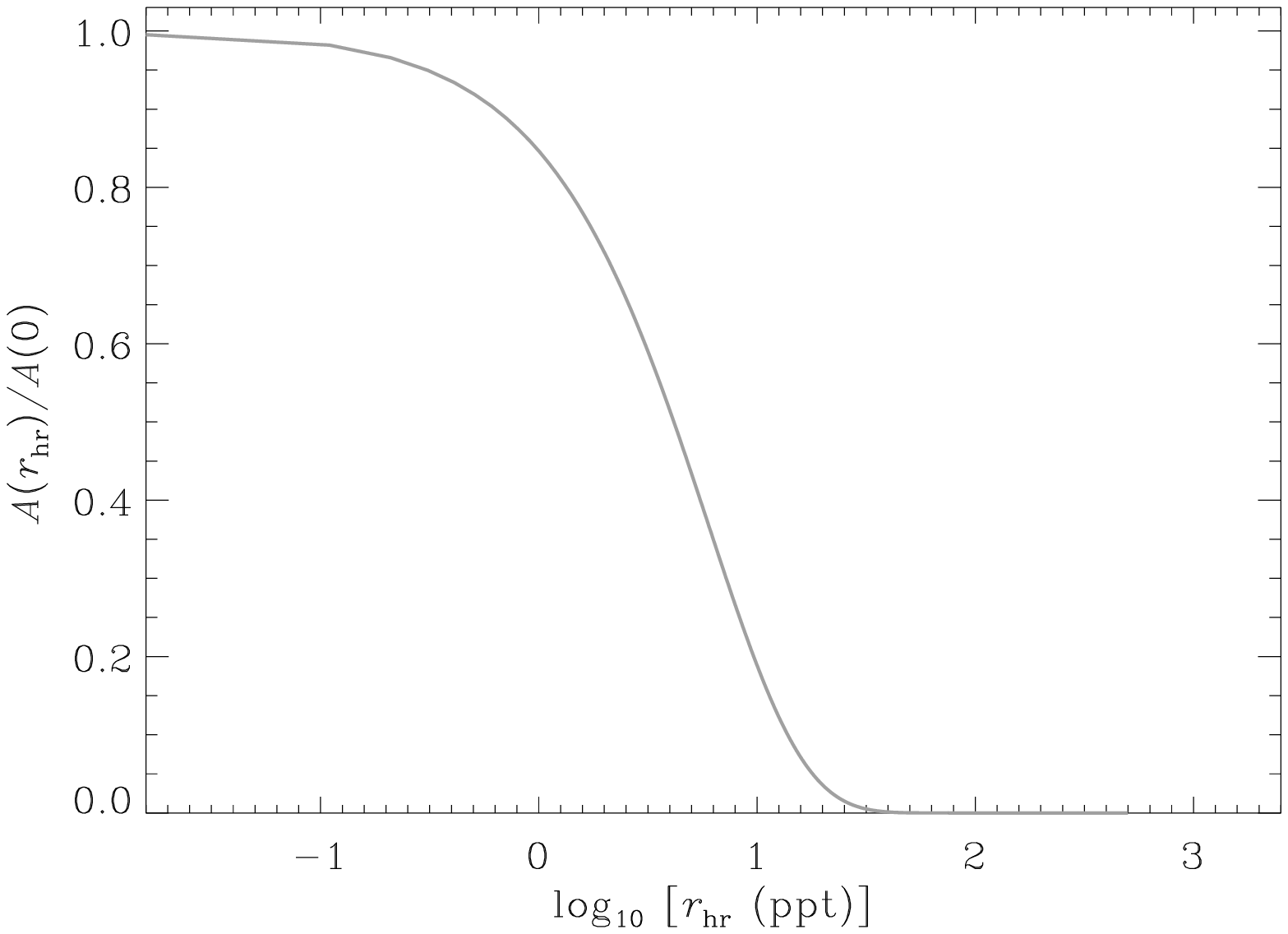}
\caption{A simple model of the suppression of mode amplitudes by
  stellar activity, with activity measured by the range parameter,
  $r_{\rm hr}$. The figure shows the expected amplitude versus $r_{\rm
  hr}$, as a fraction of the amplitude expected for zero activity
  (i.e., $r_{\rm hr}=0$).}
\label{fig:ampc}
\end{figure}

We assume that the fractional amount by which the amplitudes are
suppressed, $\delta A/A$, is a linear function in $r_{\rm hr}$. This is a
valid assumption for the solar values, but may of course be
questionable at higher levels of activity. There is also the impact of
the inclination of the star to consider (see above).  With these
caveats in mind, we have that
 \begin{equation}
 \delta A/A \simeq - \left( \frac{0.125}{1.0-0.25}\right)r_{\rm hr} \simeq
 - r_{\rm hr}/6.
 \end{equation}
Integration of the above gives the resulting suppressed amplitude,
expressed as a fraction of the amplitude expected for no activity
(i.e., zero $r_{\rm hr}$):
 \begin{equation}
 A(r_{\rm hr})/A(0) \simeq \exp \left( -r_{\rm hr}/6 \right).
 \end{equation}
Fig.~\ref{fig:ampc} plots $A(r_{\rm hr})/A(0)$ as a function of
$r_{\rm hr}$.  At a range of 20\,ppt, the prediction is that the
amplitudes are suppressed by a factor of almost 30, and it is
therefore not surprising that we see hardly any detections in the
\emph{Kepler} ensemble at this value, and none above.

%%%%%%%%%%%%%%%%%%%%%%%%%%%%%%%%%%%%%%%%%%%%%%%%%%%%%%%%%%%%%%%%%%%%%%%

%%%%%%%%%%%%%%%%%%%%%%%%%%%%%%%%%%%%%%%%%%%%%%%%%%%%%%%%%%%%%%%%%%%%%%%

The predictions in Fig.~\ref{fig:ampc} are also in agreement with the
\emph{CoRoT} results on the active G-type dwarf HD175726 reported by
Mosser et al. (2009b). Peak-to-peak variations in the lightcurve, due
to rotational modulation by spots, were found to be typically 1\,\%,
so that $r_{\rm hr} = 5\,\rm ppt$. Mosser et al. measured amplitudes
that were about 1.7-times lower than expected, based on
scaling-relation predictions that are calibrated against solar (i.e.,
low activity) values. Fig.~\ref{fig:ampc} implies that at this $r_{\rm
hr}$ amplitudes should be suppressed by a factor of about two, very
close to the factor reported by Mosser et al.

Around 100 of the \emph{Kepler} stars showing detected solar-like
oscillations will be observed for periods lasting several months up to
a few years. This will in principle allow us to further constrain the
effects of magnetic activity on the oscillation amplitudes by
measuring changes to the amplitudes, and the resulting detectability
of the modes, as activity levels vary in time. An additional 100
stars, selected by the \emph{Kepler} Science Team as possible planet
hosts, should show oscillations based on Chaplin et al. (2011b) and
will be followed for several months to years.

\acknowledgements Funding for this Discovery mission is provided by
NASA's Science Mission Directorate. The authors wish to thank the
entire \emph{Kepler} team, without whom these results would not be
possible.  We also thank all funding councils and agencies that have
supported the activities of KASC Working Group\,1, and the
International Space Science Institute (ISSI).

{\it Facilities:} \facility{The Kepler Mission}

\end{document}